# Meta-Uncertainty for Particle Image Velocimetry

Lalit K. Rajendran[1], Sayantan Bhattacharya[2], Sally P. M. Bane[1], and Pavlos P. Vlachos[2*]

[1] Purdue University, School of Aeronautics and Astronautics, West Lafayette, USA.

[2] Purdue University, School of Mechanical Engineering, West Lafayette, USA.

*pvlachos@purdue.edu

## Abstract

Uncertainty quantification for Particle Image Velocimetry (PIV) is critical for comparing experimentally measured flow fields with Computational Fluid Dynamics (CFD) results, and model design and validation. However, PIV features a complex measurement chain with coupled, non-linear error sources, and quantifying the uncertainty is challenging. Multiple assessments show that none of the current methods can reliably measure the actual uncertainty across a wide range of experiments, and estimates can vary. Because the current methods differ in assumptions regarding the measurement process and calculation procedures, it is not clear which method is best to use for an experiment where the error distribution is unknown.

To address this issue, we propose a method to estimate an uncertainty method's sensitivity and reliability, termed the Meta-Uncertainty. The novel approach is automated, local, and instantaneous, and based on perturbation of the recorded particle images. We developed an image perturbation scheme based on adding random unmatched particles to the interrogation window pair considering the signal-to-noise (SNR) of the correlation plane. Each uncertainty scheme's response to several trials of random particle addition is used to estimate a reliability metric, defined as the rate of change of the inter-quartile range (IQR) of the uncertainties with increasing levels of particle addition. We also propose applying the meta-uncertainty as a weighting metric to combine uncertainty estimates from individual schemes, based on ideas from the consensus forecasting literature. We use planar and stereo PIV measurements across a range of canonical flows to assess the performance of the uncertainty schemes. Further, a novel method is introduced to assess an uncertainty scheme's performance based on a quantile comparison of the error and uncertainty distributions, generalizing the current method of comparing the RMS of the two distributions. The results show that the combined uncertainty method outperforms the individual methods, and this work establishes the meta-uncertainty as a useful reliability assessment tool for PIV uncertainty quantification.

# Nomenclature

| | | | |
|---|---|---|---|
| $\epsilon$ | Error | $n$ | Individual method index |
| $\Delta x$ | Displacement along the x-direction | $N$ | Number of individual methods |
| $\sigma$ | Standard deviation | $\mathcal{N}$ | Gaussian/normal distribution |
| $\sigma_{\Delta x}$ | Uncertainty in x displacement | $U$ | Uncertainty |
| $\sigma_{rat}$ | Ratio of uncertainties of perturbed and original images | $x$ | Horizontal co-ordinate |
| $f$ | Probability density function (PDF) | $y$ | Vertical co-ordinate |
| $m$ | Rate of change of uncertainty ratio with particle addition | $w$ | Weight |

# 1  Introduction

Over the past decade, there has been increasing effort in Particle Image Velocimetry (PIV) to develop a-posteriori uncertainty quantification methodologies for local and instantaneous displacement measurements [1]. These efforts aim to provide uncertainties to PIV measurements that can be used for comparison to CFD results and for model design and validation. Parallel efforts included methods for propagating these uncertainties to derived quantities such as turbulence statistics [2], velocity derivatives [3], and pressure [4, 5], as well as to stereo PIV [6] and volumetric PTV [7] measurements. There has also been recent progress in uncertainty quantification in Background Oriented Schlieren (BOS), an image-based density measurement technique. Developments include estimation of displacement uncertainty from dot tracking based processing [8, 9], propagation of both tracking/cross-correlation based displacement uncertainties through the density integration chain [10], and utilizing the uncertainties to improve the density integration process by weighted least squares minimization [11].

The measurement uncertainty of a quantity (e.g., velocity in PIV) represents the interval expected to contain the true value. This uncertainty depends on all the factors in the overall measurement chain. Since PIV involves a complex measurement chain from image recording through processing and post-processing, the final measurement can suffer from a multitude of error sources such as particle size, seeding density, shear, noise, out-of-plane motion, and processing algorithms, to name a few [12]. These error sources can combine in a coupled and non-linear manner to affect the final measurement uncertainty, and also depend on the final quantity of interest, whether the displacement, shear, pressure, or density. Even just for displacement uncertainty, while there have been many methods proposed in the literature, none perform well under all situations.

PIV displacement uncertainty methods are commonly classified into *direct* and *indirect* methods. Indirect methods predict the displacement uncertainty by calibrating the variation of uncertainty to various image parameters (such as particle size, density, shear, noise) and signal-to-noise ratio metrics of the cross-correlation plane (such as the Peak to Peak Ratio (PPR), Mutual Information (MI) and others [13–15]). Monte-Carlo simulations with synthetic images are used to obtain the calibration [13–16]. The performance of all indirect methods relies on the calibration process, which must be accurate and reflect all possible experimental scenarios in a typical measurement.

Direct methods estimate the uncertainty directly based on image or correlation plane properties without calibration. Presently, three direct methods are available to estimate the displacement uncertainty—Image Matching (IM) [17], Correlation Statistics (CS) [18], and Moment of Correlation (MC) [19]. In brief, IM or particle disparity (PD) proposed by Sciacchitano *et al*. [17] estimates the uncertainty in the displacement using a statistical analysis of the disparity between the measured positions of particles in the two frames after a converged iterative deformation interrogation procedure. The performance of this method is sensitive to the accuracy of the particle position estimation and deteriorates with increasing seeding density, noise, and out-of-plane motion. CS, proposed by Wieneke [18], estimates the uncertainty again using the image disparity but at a pixel level. The correlation peak's asymmetry at the end of a converged window deformation procedure is used to measure the correlation error, and propagating the standard deviation of this error through the sub-pixel estimator provides the uncertainty. The CS method relies on the correlation plane statistics and performs better at higher seeding densities and larger interrogation windows. MC, proposed by Bhattacharya *et al*. [19], predicts the uncertainty by estimating the second-order moment of the PDF of displacements contributing to the cross-correlation plane. The PDF is estimated as the generalized cross-correlation (GCC) from the inverse Fourier transform of the phase of the complex cross-correlation plane [20–22], followed by Gaussian filtering, gradient correction, and scaling by the effective number of particles contributing to the cross-correlation. This method also works better with high seeding densities and large interrogation windows, and small interrogation windows can lead to an over-prediction of the uncertainty.

However, multiple previous works show that none of the PIV uncertainty quantification methods perform well under all situations [23, 24]. While the direct methods are sensitive to elemental error sources [23], they can under-predict the random error [24]. In addition, direct methods can predict different uncertainties for the same flow field [10, 19], and as a result, no PIV uncertainty method is universally consistent and robust. Further, it is often impossible to choose the correct estimate in an experiment because the actual random error is unknown, and these potentially incorrect estimates in displacement uncertainty can propagate to derived quantities with detrimental implications for further analysis. Therefore, it is not clear which method to use for an experiment where the error is unknown.

A similar problem also exists in the consensus forecasting literature when assessing the risk/reliability associated with competing models that predict a future quantity based on incomplete information in the present [25–27]. In these applications, the variance of the fluctuations of each model prediction provides the "risk"/"volatility". In this work, we adapt this idea to the problem of PIV uncertainty quantification and develop a method to estimate the robustness/sensitivity of each uncertainty method in a local, instantaneous, and automated manner. We base the method on perturbing the particle images in an interrogation window pair and assessing the variation of the uncertainty estimates to this perturbation. The perturbation should be large enough to provide a variation of the displacement uncertainty *without* significantly affecting the cross-correlation plane and displacement estimate on which we are trying to calculate the uncertainty bounds. We assess this using signal-to-noise ratio metrics of the cross-correlation, such as the peak ratio and mutual information. By repeating the random particle addition over several trials and over different addition amounts, we quantify the response of each uncertainty scheme to the image perturbation. Finally, we use descriptive statistics of the distribution, such as the inter-quartile range (IQR) and the rate of change of the IQR with increasing particle addition, to estimate a reliability metric (the *meta-uncertainty)* for each uncertainty scheme. A broader distribution of uncertainty estimates and a higher meta-uncertainty will characterize schemes that are more sensitive to the perturbations.

Finally, we apply the meta-uncertainty to develop a new uncertainty quantification scheme for PIV that combines the estimates from the individual schemes weighted by the inverse of their meta uncertainty. Similar to consensus forecasting, where the aim is to combine estimates from different models based on a meta-analysis of the individual models [26–28], this approach aims to fuse the prediction from multiple uncertainty models into a new, more robust, and reliable estimate. The hypothesis is that different models utilize different aspects of the information associated with the measurement, and therefore their combination provides a better estimate than each individual model. In the present context, the forecast quantity is the displacement uncertainty, the individual models are the uncertainty quantification methods, and the meta-uncertainty provides the weights for each model. While the proposed framework is general and can apply to many individual uncertainty schemes, here, we will consider only the three *direct* displacement uncertainty schemes—Image Matching (IM) [17], Correlation Statistics (CS) [18], and Moment of Correlation (MC) [19]. We assess the performance of the meta-uncertainty estimation method and the combined uncertainty scheme with synthetic and experimental planar and stereo PIV images.

## 2  Methodology

Figure 1 shows the overall meta-uncertainty based combination method which consists of three major steps: 1) the estimation of the meta-uncertainty for each uncertainty method, 2) calculation of weights based on the response function, and 3) calculating the combined uncertainty. The following sections detail the procedure for each step.

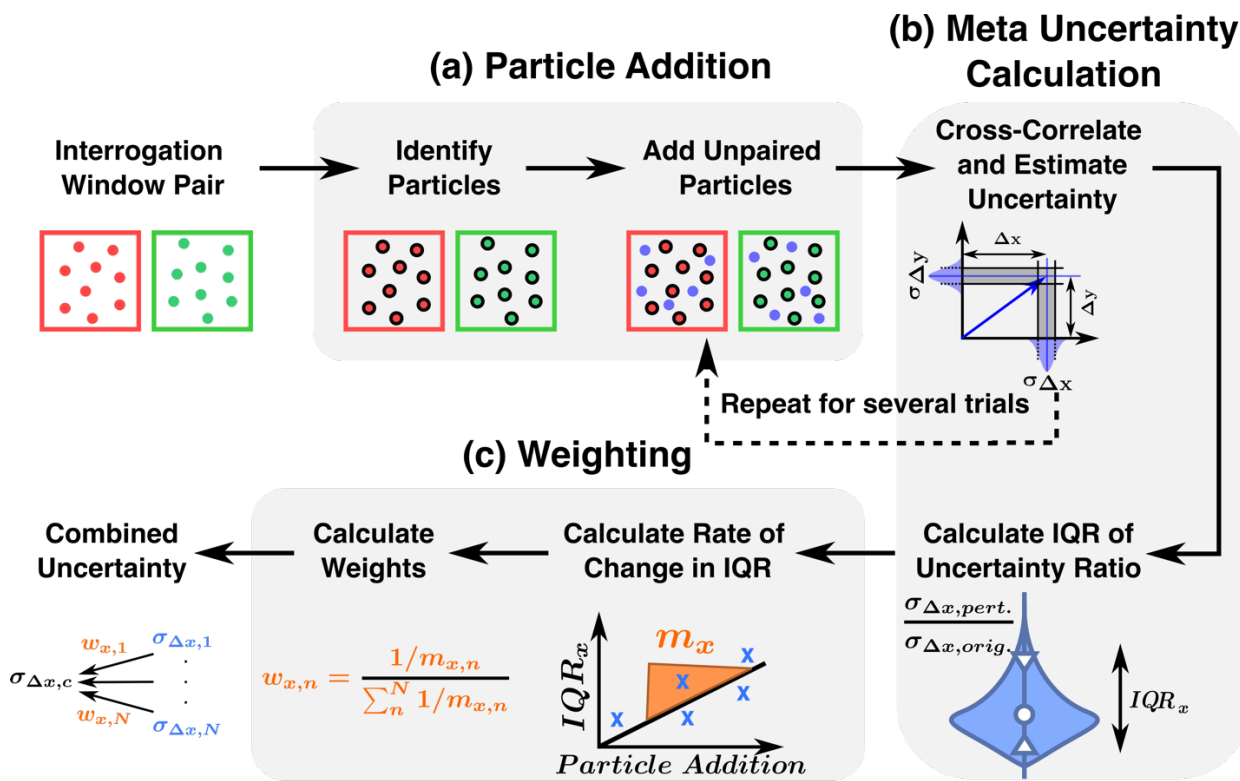

**Figure 1.** Illustration of the meta-uncertainty based combination methodology.

The meta-uncertainty is based on the uncertainty scheme's PDF and describes its response to a perturbation in the input intensity distribution. Since the PIV uncertainty methods rarely have a closed-form expression, we estimate the PDF using a Monte-Carlo simulation procedure. Further, to estimate the *true/parent* PDF requires the knowledge of all inputs—a set of all possible PIV images—which is not possible. Therefore, we perturb the intensity distributions of the interrogation window pair for several trials to generate a local population of particle image pairs and estimate the corresponding uncertainty.

For the image perturbation procedure to be valid, it must be able to provide a variation of the uncertainty estimates without an appreciable change in the underlying signal-to-noise ratio (SNR) metrics of the cross-correlation estimator (such as the Peak to Peak Ratio (PPR), Mutual Information (MI) and others [13–15]) whose uncertainty we are trying to estimate. There are several potential methods to perturb the particle images. We adopt a method of adding random and unpaired particles to the interrogation window pair. Analysis with synthetic and experimental images showed that this method best accomplished perturbing the images with a negligible change in the signal-to-noise ratio metrics.

*a) Particle Perturbation*

To perform the perturbation, we first identify all the particles on the image using identification and centroid-estimation methods commonly used in Particle Tracking Velocimetry [29, 30]. Following this, we add a set of unpaired particles to the interrogation window pair as shown in Figure 1 (a), with the number of unpaired particles specified as a fraction of the seeding density, and the peak intensity and diameter set to be the average of the already identified particles. Figure 2 shows a sample particle image pair with the perturbation.

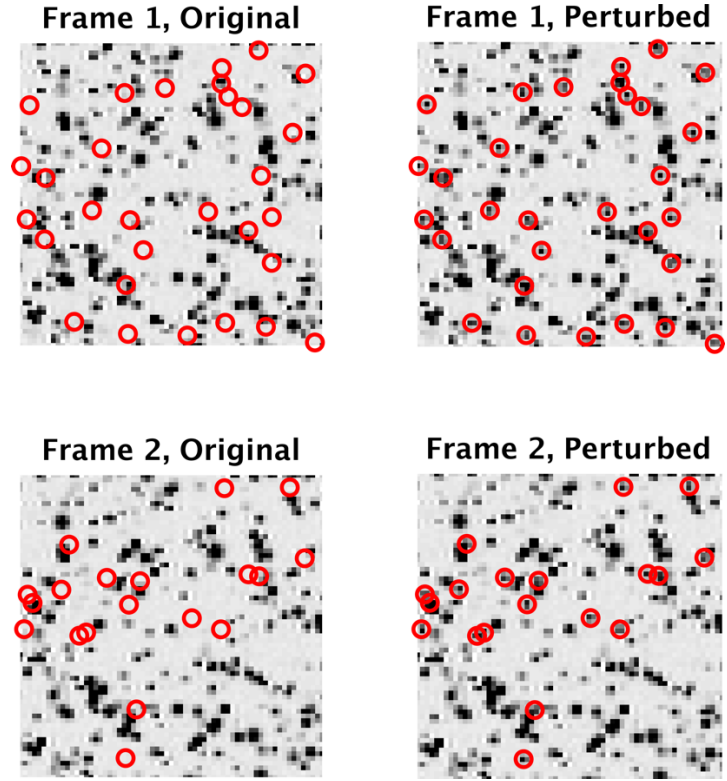

**Figure 2.** Example of a perturbed image pair, with the red circles indicating the location of the added particles.

*b) Meta uncertainty calculation*

The perturbed window pair is then cross-correlated, and the uncertainty is estimated using all three individual methods as shown in Figure 1 (b). We repeat this procedure for several trials to build a PDF of the *ratio* of the perturbed to original uncertainties for each estimator ($\sigma_{rat,\Delta x,n} = \sigma_{\Delta x,n,pert.}/\sigma_{\Delta x,n,orig.}$). Figure 3 shows sample distributions of the uncertainty schemes and statistics such as the median and quartiles**Figure 3**. Each level of particle addition results in a *distribution* of uncertainties due to the perturbation. These distributions become wider with increasing level of particle addition at a different rate of increase for each method. The width of the distribution represents the sensitivity of each scheme to particle perturbation, and therefore a scheme with a wider PDF is less reliable compared to a scheme with a narrower PDF. However, the response and relative sensitivity of each scheme will vary with the local image and flow conditions.

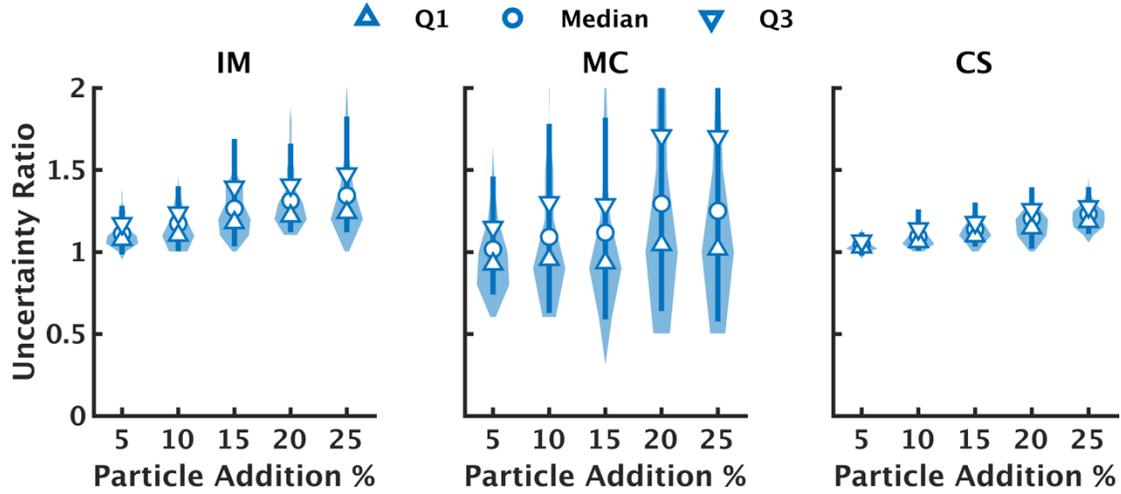

**Figure 3.** Effect of particle addition on the PDF of the ratio of resampled to original uncertainties.

*c) Weight calculation*

We calculate the IQR for each particle perturbation level from the distributions, and the rate of change of this IQR with particle perturbation (from a linear regression as shown in Figure 1 (c)) provides the weight for each scheme. The IQR is used in place of the RMS because it is less sensitive to outliers. Equation (1) provides the weight of the x component(1), with a similar equation for the *y* component. The slope (and the weight) calculation depends on the number of points used for the regression and the individual scheme's response. Figure 4 shows a sample result for the IQR variation with five particle addition levels and the corresponding weights**Figure 4**. These results are consistent with Figure 3 with MC showing the highest rate of increase and therefore assigned the lowest weight, with CS showing the lowest rate of increase and therefore assigned the highest weight. This weight is essentially a local and instantaneous reliability assessment metric for each uncertainty scheme, and the proposed method allows for an automated way to estimate this metric for arbitrary particle images. The relative weights for each scheme can vary across grid points within the same flow field, and across flow-fields.

$$w_x = \left|\left(\frac{\Delta\, IQR_x}{\Delta\, particle\, \%}\right)\right|^{-1} \tag{1}$$

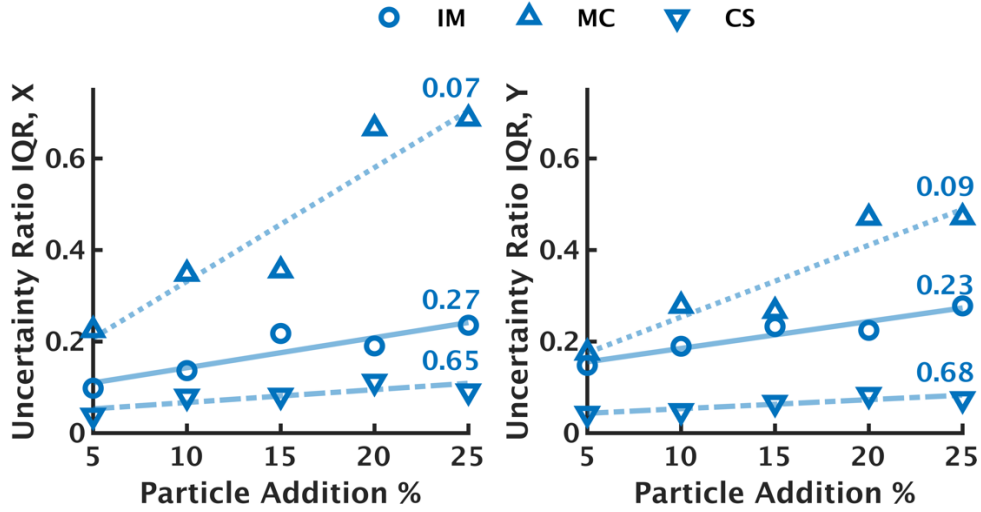

**Figure 4.** Variation of the IQR of uncertainty ratios with particle addition percentage for a grid point, with the corresponding weights obtained from the straight line fit.

*d) Combined uncertainty calculation*

Finally, we calculate the combined uncertainty as the weighted average of the individual uncertainty schemes as shown in Figure 1 (d),

$$\sigma_{x,comb} = \sum_{n=1}^{N} w_{x,n} \sigma_{x,n} \qquad (2)$$

for the x-component where $\sigma_x$ represents the individual uncertainty, $w_x$ represents the corresponding weights, $n$ represents the subscript for each method, and $N$ represents the total number of methods (here, $N = 3$). In the next section, we will assess the method's performance with synthetic and experimental images from planar and stereo PIV experiments.

## 3  Results

### 3.1  Planar PIV

Planar PIV measurements from several canonical flows are used to test the uncertainty quantification methods over a wide variation of image and flow conditions. The datasets used are: a turbulent boundary layer (PIV Challenge 2003B) [31], a laminar separation bubble (PIV Challenge 2005B) [32], laminar stagnation flow [33], a vortex ring (fourth PIV Challenge) [34], and the unsteady inviscid core of a jet [23]. For each dataset, we processed the images with two processing routines (WS1 and WS2 as listed in Table 1) to provide a further variation in the testing, using the open-source PIV code PRANA [35, 36]. Figure 5 shows the displacement contours from all flow fields, and foFigure **5** each case, the error analysis used a true solution, based on details from the respective publications.

**Table 1.** Summary of processing parameters for all datasets.

|  | **Turbulent Boundary Layer (TBL)** | **Laminar separation bubble (LSB)** | **Stagnation flow (SF)** | **Vortex ring (VR)** | **Jet flow (JF)** |
|---|---|---|---|---|---|
| **WS 1 (% overlap, No. of passes)** | 64 × 64 (75%, 2) (87.5%, 2) | 64 × 64 (75%, 4) | 64 × 64 (75%, 4) | 64 × 64 (75%, 1) (87.5%, 3) | 32 × 32 (87.5%, 4) |
| **WS 2 (% overlap, No. of passes)** | 64 × 64 (87.5%, 1) 32 × 32 (75%, 3) | 64 × 64 (75%, 1) 32 × 32 (50%, 3) | 64 × 64 (75%, 1) 32 × 32 (50%, 3) | 64 × 64 (87.5%, 1) 32 × 32 (75%, 3) | 32 × 32 (75%, 1) 16 × 16 (75%, 3) |

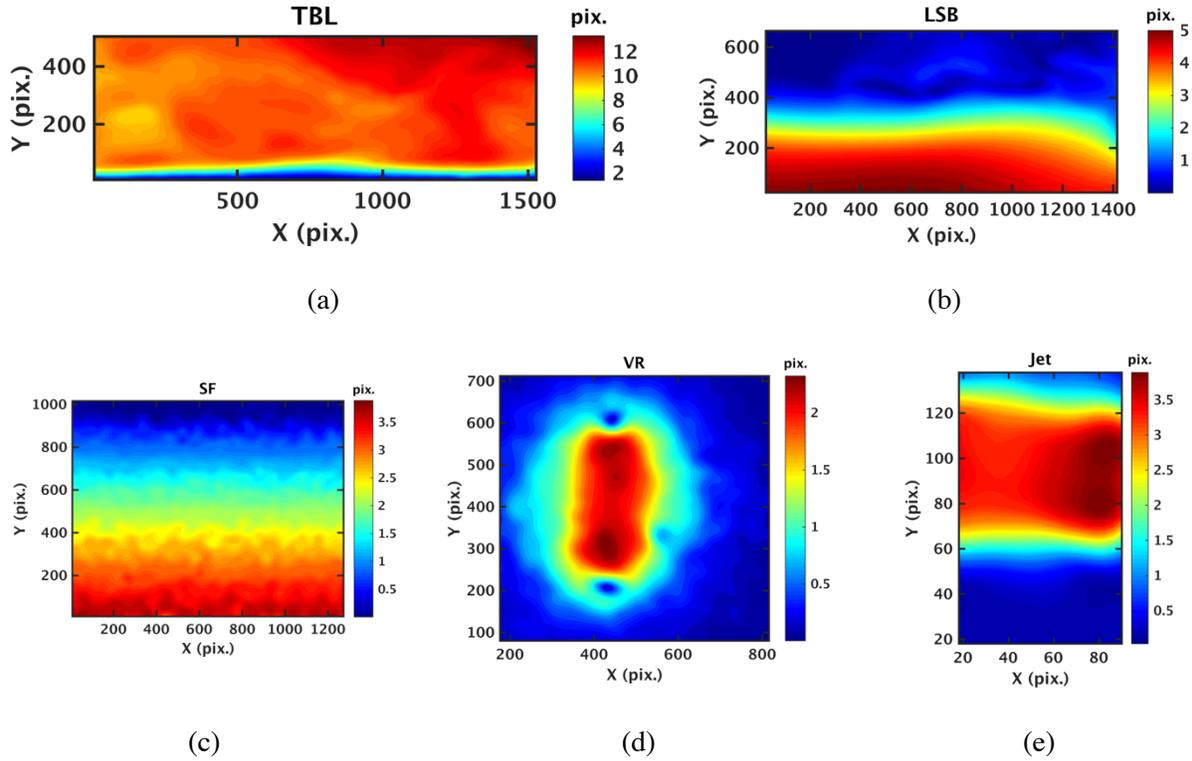

**Figure 5.** Datasets used for assessment on planar PIV images. (a) Turbulent boundary layer [31], (b) laminar separation bubble [32], (c) laminar stagnation flow [33], (d) vortex ring [34], and (e) jet flow [23].

We test the combination framework shown on the datasets using a Monte-Carlo simulation by randomly choosing a dataset and a processing setting, a snapshot within the dataset, and a grid point within the snapshot. For the chosen grid point, particle addition is used to calculate the meta-uncertainty, and the individual uncertainties are combined using a weighted average using the procedure shown in Figure 1. Particle addition was performed for five levels of the seeding density from 5 to 25% and over 100 trials for each perturbation level (as shown in Figure 3 and Figure 4). This procedure is repeated for 1000 grid points for each dataset and processing setting, and the errors and uncertainties across all datasets (10,000 grid points in total) are merged to calculate the corresponding statistics. This section discusses the merged statistics, and the Appendix contains the individual dataset results.

Figure 6 shows the PDFs of the error and uncertainty distributions **Figure 6** in the form of violin plots, with the individual schemes in blue, the error in black, and the combined uncertainty scheme in orange. Also shown are the statistics such as the median (circles), quartiles (triangles), and the root mean square (RMS)(straight line) of each distribution. Sciacchitano et al. [23] showed that when the error distribution for each grid point is modeled as a zero-mean Gaussian random variable with the standard deviation representing the local uncertainty, the RMS of a mixture of these error distributions should match the RMS of the corresponding uncertainty distributions. For these results, the RMS of the error distribution is 0.08 pix., with the combined uncertainty scheme predicting an RMS of 0.07 pix., while the RMS of the individual uncertainty schemes being 0.05 pix. for IM, 0.06 pix. for MC, and 0.1 pix. for CS. Therefore, the combined scheme provides the best estimate of the RMS error and can compensate for the under-prediction by IM and MC and the over-prediction by CS. However, there is still a 0.01 pix. discrepancy between the RMS estimates of the error and the combined scheme, indicating that there is room for further improvement of the method.

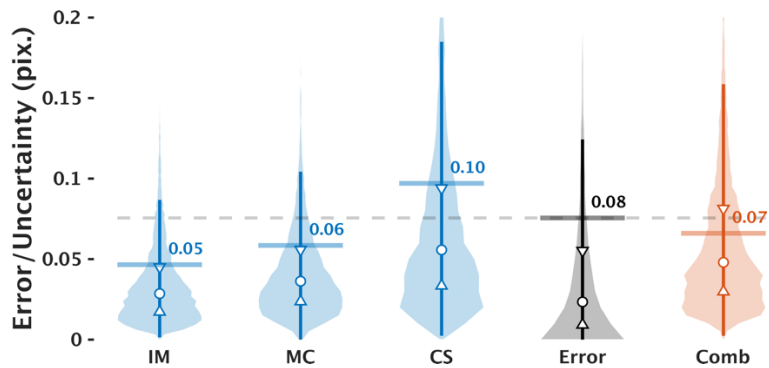

**Figure 6.** PDFs of the error and the individual and combined uncertainty estimates for consolidated results from all datasets.

The PDF of weights assigned to each individual scheme are shown in Figure 7 with the individual schemes in blue, and the black dashed line represents the case when all schemes are equally weighted. CS is assigned the highest weights for most cases, followed by IM, and then by MC, consistent with the sample result shown in Figure 4. However, even though CS is assigned a higher weight and over-predicts the RMS of the error, the combined effect of IM and MC, which under-predict the error, brings down the RMS of the combined scheme to be close to the RMS of the error. This highlights the advantage of the meta-model as even if one uncertainty scheme (CS) is more robust to perturbations in the correlation plane SNR and hence has a higher weight, the method can compensate with the weighting of other schemes. Finally, the weight distributions are broad, denoting that there are several grid points for which CS could be assigned a lower weight than either IM and MC. Therefore, the meta-uncertainty calculation is also able to capture the variation in the image and flow conditions across the datasets.

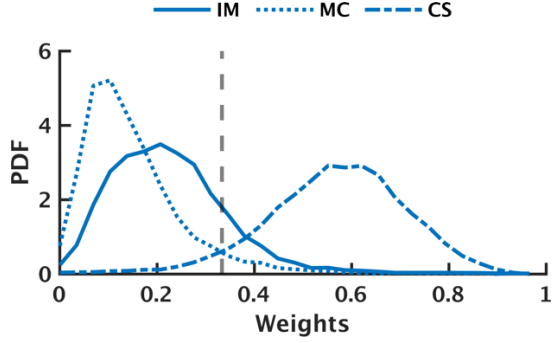

**Figure 7.** Distribution of weights assigned to the individual uncertainty schemes across all the datasets. The grey dashed line represents a case with equal weighting.

We also introduce a new method to compare the error and uncertainty distributions, based on a quantile-quantile comparison of the error and uncertainty distributions. This is a generalization of the method proposed by Sciacchitano et al. [23] for comparing the RMS to address the sensitivity of the RMS calculations to outliers. Consider a set of error measurements $\epsilon_i$, where $i$ represents the grid point under consideration with each error drawn from a corresponding distribution $f_{\epsilon_i}$. Also, let the distribution of all the error measurements be $f_\epsilon$. For each error measurement, we have an estimate of the uncertainty $U_{n,i}$ where $n$ represents an uncertainty method. This uncertainty measurement represents the standard deviation of the error distribution $f_{\hat{\epsilon}_{n,i}}$, with $\hat{\epsilon}_{n,i}$ representing the *estimated* error. Finally, we can also define a combined distribution of each uncertainty scheme's estimated errors as $f_{\hat{\epsilon}_n}$. Our aim then is to compare the distributions of the *true* error distribution, $f_\epsilon$, with that of the *estimated* error distribution, $f_{\hat{\epsilon}_n}$, for each uncertainty scheme. To enable this comparison, we need a model for the individual, estimated, error distributions $f_{\hat{\epsilon}_{n,i}}$. Following the analysis of Sciacchitano et al. [23], if we assume this distribution to be a zero-mean Gaussian random variable, ($f_{\hat{\epsilon}_{n,i}} = \mathcal{N}(0, U_{n,i})$) then the *overall* distribution of the *estimated* error becomes the sum of these individual distributions,

$$f_{\hat{\epsilon}_n} = \sum_i f_{\hat{\epsilon}_{n,i}} = \sum_i \mathcal{N}(0, U_{n,i}) \ . \tag{3}$$

If the uncertainty estimate $U_{n,i}$ is correct, then the RMS of the above distribution must equal that of the error distribution, consistent with the previous result of Sciacchitano et al. [23]. Therefore, in this work, we compare the *distributions* instead of the RMS values for a more rigorous comparison and to reduce the effect of outliers.

The following procedure is used to estimate $f_{\hat{\epsilon}_n}$. For each grid point and uncertainty method $U_{n,i}$, we draw several (here 1000) random values of $\hat{\epsilon}_{n,i}$ from the corresponding normal distribution. These estimated error values from all grid points provide a pdf of the estimated error distribution $f_{\hat{\epsilon}_n}$. Then the true and estimated error distribution are compared using a quantile-quantile plot.

The results are shown in Figure 8, where the x-axis represents the quantiles of $f_\epsilon$ and the y-axis represents the quantiles of $f_{\hat{\epsilon}_n}$, with each curve corresponding to an uncertainty scheme, and the black line representing the 1:1 variation. The orange curve corresponding to the combined scheme is overall closest to the black line, showing that the combined uncertainty scheme best approximates the true error distribution**Figure 6**.

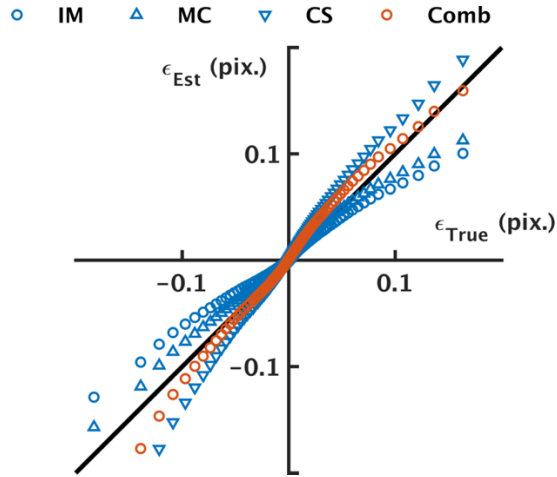

**Figure 8.** Quantile-quantile comparison of the true and estimated error distributions.

To complement the statistical analysis, we investigate the variation of the RMS error and uncertainty as functions of the element error sources such as fractional displacement and shear [16, 37]. To perform the comparison, we bin the errors and uncertainties based on their corresponding values of the displacement and shear, as estimated from the *true solution*. Then we calculate the bin-wise RMS of the error and uncertainties to caFigure **9** the variation of these statistics with the elemental error sources. Figure 9 shows these results, along with the number of measurements corresponding to each bin. The results show that 1) the errors/uncertainties increase with fractional displacement and (to a lesser extent) with shear, which is consistent with PIV theory [38–40], and 2) the combined scheme provides an RMS uncertainty that is, on average, the closest to the RMS error. However, MC performs better for low values of the velocity gradients, since the large uncertainties predicted by CS for these measurements shift the combined estimates upward.

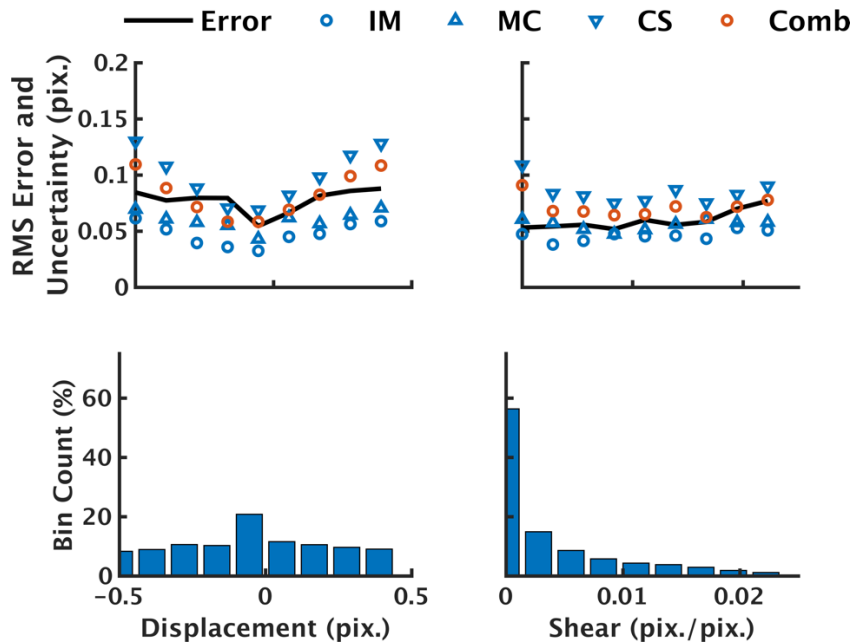

**Figure 9.** Variation of the RMS error and uncertainty as a function of elemental error sources such as displacement and shear, along with the corresponding bin count.

In summary, the analysis on planar PIV datasets showed that the combined uncertainty scheme based on the meta-uncertainty better represented the error distribution in terms of the RMS, quantiles, and the effect of error sources such as fractional displacement and shear. In the next section, we demonstrate the performance of the method for Stereo PIV images of a vortex ring.

## 3.2 Stereo PIV

The performance of the meta-uncertainty-based framework is also tested with Stereo PIV images by utilizing the uncertainty quantification methodology introduced by Bhattacharya et al. [6]. The method propagates the planar PIV uncertainties for each camera through the stereo-reconstruction process, accounting for uncertainties in the mapping function coefficients from the self-calibration procedure [41]. The analysis here uses the vortex ring dataset from Case E of the 4$^{th}$ PIV Challenge [34] (center and left cameras), similar to Bhattacharya et al. [6]. Figure 10 shows displacement contours for the three displacement components, with 50 snapshots used for the analysis.

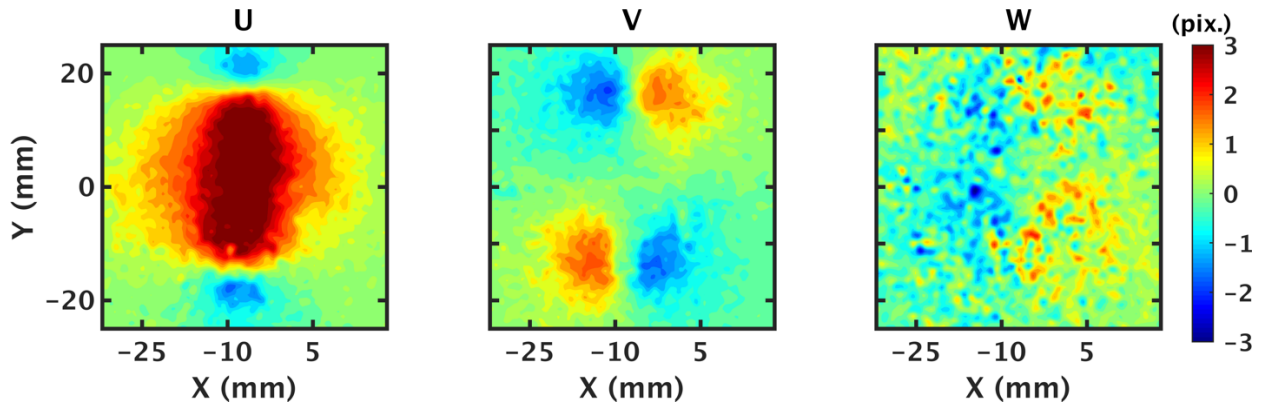

**Figure 10.** Spatial variation of displacement components for the stereo PIV dataset.

We assess the uncertainty schemes using the Monte-Carlo procedure detailed before, with the additional step of propagating the perturbed planar uncertainties through the stereo-reconstruction procedure to calculate the corresponding *stereo* uncertainties, the IQR and the weights. Therefore, the full measurement chain was used to calculate the meta and combined uncertainties.

Figure 11 shows the distribution of weights, errors, and uncertainties, and Figure 11**Figure 11Figure 11**(a) is consistent with the planar results, with CS being assigned the highest weight, followed by IM and MC. Further, the relative distribution of the weights is nearly identical for the three displacement components. From the error and uncertainty distributions shown in Figure 11(b), we see that the RMS of the combined uncertainty method is again the closest to the RMS error for the in-plane component U and V, similar to the planar data, and slightly over-predicts the RMS for the out-of-plane component W. This over-prediction (~ 0.1 pix.) is because of the over-prediction in the CS estimates of the uncertainty.

Finally, the quantile-quantile plots of the true and estimated error distributions in Figure 11(c) show that the combined uncertainty best approximates the true error distribution for the in-plane displacement components, while IM and MC perform better for the out-of-plane components. The deviation of the combined uncertainty closely follows that of the CS curve because of the high weights assigned to CS. Therefore, in situations without an obvious choice for the best individual scheme, the combined scheme offers minor performance improvement. However, the performance of the individual schemesvaries for the vast majority of the experiments, and when the error distribution is not available, it is impossible to guess the best method. Therefore, the combined method offers the most robust estimate of the uncertainty for a general experiment without a true solution. Overall, these results establish that the meta-uncertainty based combination framework also performs well for Stereo PIV measurements.

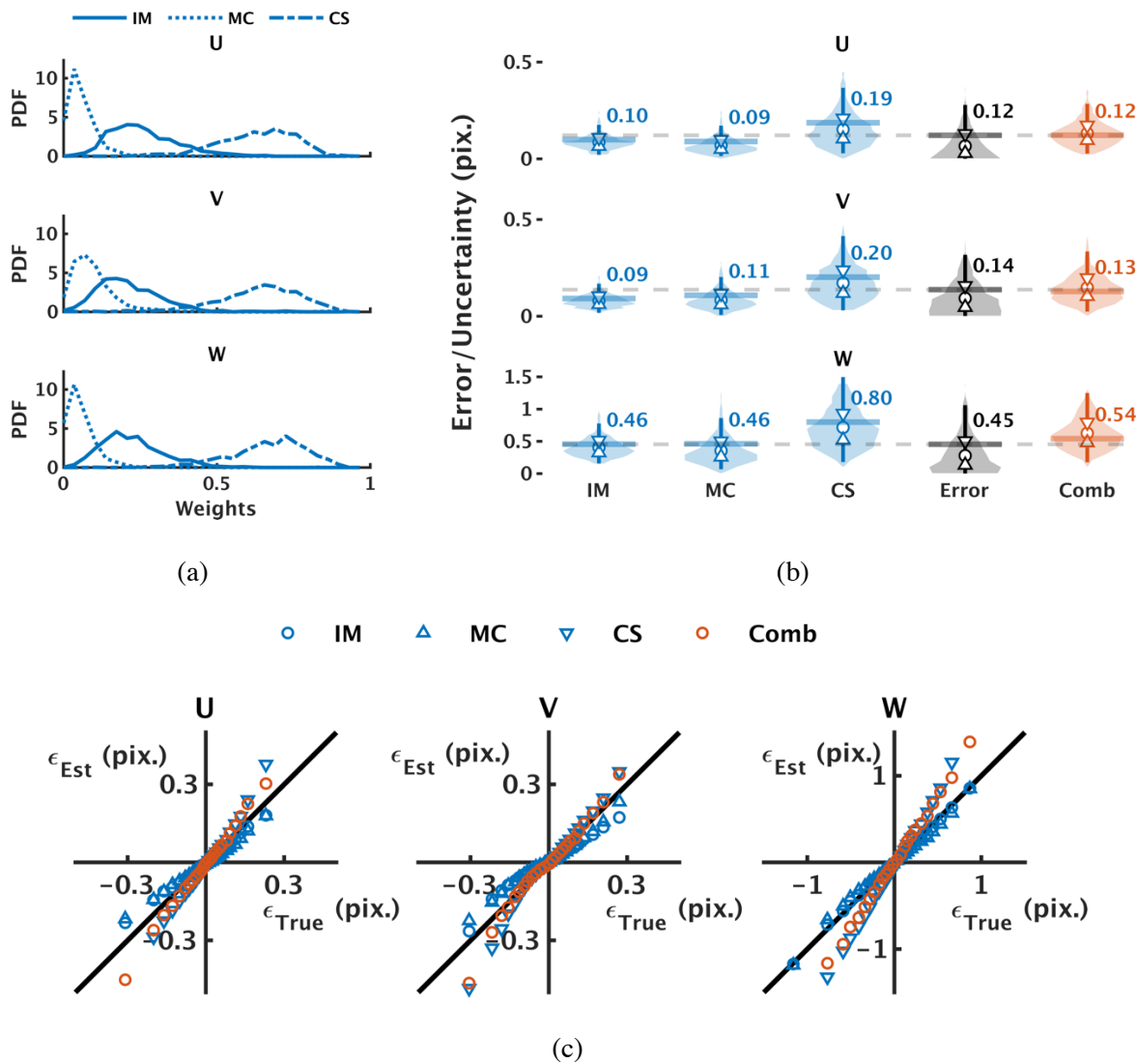

**Figure 11.** Results of applying the meta-uncertainty model to stereo PIV images. (a) PDF of weights, (b) rror and uncertainty distributions, and (c) Quantile-quantile comparison of the true and estimated error.

## 4 Summary and Conclusions

This work introduced a reliability metric of PIV uncertainty quantification methods termed the *meta-uncertainty* and an automated, local, and instantaneous method for its estimation. The meta-uncertainty describes the sensitivity of an uncertainty quantification method to perturbation in the input images, with a more sensitive scheme possessing a higher meta-uncertainty and lower reliability. Random/unpaired particles are added to perturb the images and estimates the uncertainty using each method over several trials and different particle addition levels. The PDF of the uncertainty provides a statistical measure of the response function, and the rate of change of the inter-quartile range (IQR) of the individual uncertainty schemes with particle addition provides the reliability metric.

In addition, this work also introduced a framework for combining individual uncertainty estimates based on the meta-uncertainty, similar to consensus forecasting. Since the uncertainty estimation methods differ in their use of information regarding the displacement estimation process, we hypothesized that combining the individual estimates should provide a better estimate of the uncertainty. The individual estimates were combined using a weighted average, with the weights based on the inverse of the rate of change of IQR, with a more sensitive/less reliable scheme assigned a lower weight.

Both the meta-uncertainty estimation and the combination framework were tested with the direct uncertainty methods - Image Matching (IM), Moment of Correlation (MC), and Correlation Statistics (CS) - with planar and stereo PIV images of several canonical flows, which offer a range of error and uncertainty sources. The planar PIV dataset included a turbulent boundary layer, laminar separation bubble, laminar stagnation flow, vortex ring, and jet flow, two processing settings for each flow field, and the stereo PIV dataset used was a vortex ring. We calculated the individual and combined uncertainties for grid points randomly sampled from these datasets, and results showed that the combined uncertainty best represented the true error distribution in terms of the overall RMS and the variation of RMS with error sources such as displacement and shear. Further, a new method was introduced to compare the error and uncertainty estimates by generalizing the RMS comparison method of Sciacchitano et al. [23] to quantiles of the error and uncertainty distribution for a more rigorous comparison that is less sensitive to outliers. These results also showed that the error distribution based on the combined method predicts the true error distribution better than the individual uncertainty methods. For the stereo PIV dataset, the meta-uncertainty based combined method showed the best performance for the in-plane components, with a slight over-prediction in the RMS error for the out-of-plane component. However, since an individual uncertainty schemes performance varies significantly for different experiments, the combined method likely provides the best potential estimate of the uncertainty for a general experiment.

The major limitation of the method is the computational cost involved in estimating the meta-uncertainty with approximately 1000 perturbation trials performed for each grid point and uncertainty scheme. Therefore, future work could reduce this computational cost by developing approximate theoretical models of the individual schemes' response functions to accelerate the computations. Machine learning based neural-network models can also improve the combination framework. Finally, the meta-uncertainty can also improve the individual schemes themselves by analyzing their response to particle perturbations. In conclusion, this paper establishes the meta-uncertainty as a useful reliability assessment tool for PIV uncertainty quantification and the combination framework as a successful estimator of the uncertainty for cross-correlation PIV processing, with potential applications to uncertainty propagation, de-noising, and other post-processing routines.

## 5  Acknowledgment

The US Department of Energy, Office of Science, Office of Fusion Energy Sciences supported this work under Award Number DE-SC0018156, and NSF provided support through 1706474.

# 6 References

1. Sciacchitano A (2019) Uncertainty quantification in particle image velocimetry. *Measurement Science and Technology*, 30(9):092001. https://doi.org/10.1088/1361-6501/ab1db8

2. Wilson BM, Smith BL (2013) Uncertainty on PIV mean and fluctuating velocity due to bias and random errors. *Measurement Science and Technology*, 24(3):035302. https://doi.org/10.1088/0957-0233/24/3/035302

3. Sciacchitano A, Wieneke B (2016) PIV uncertainty propagation. *Measurement Science and Technology*, 27(8):084006. https://doi.org/10.1088/0957-0233/27/8/084006

4. Azijli I, Sciacchitano A, Ragni D, Palha A, Dwight RP (2016) A posteriori uncertainty quantification of PIV-based pressure data. *Experiments in Fluids*, 57(5):72. https://doi.org/10.1007/s00348-016-2159-z

5. Zhang J, Bhattacharya S, Vlachos PP (2020) Using uncertainty to improve pressure field reconstruction from PIV/PTV flow measurements. *Experiments in Fluids*, 61(6):131. https://doi.org/10.1007/s00348-020-02974-y

6. Bhattacharya S, Charonko JJ, Vlachos PP (2016) Stereo-particle image velocimetry uncertainty quantification. *Measurement Science and Technology*, 28(1):015301. https://doi.org/10.1088/1361-6501/28/1/015301

7. Bhattacharya S, Vlachos PP (2020) Volumetric particle tracking velocimetry (PTV) uncertainty quantification. *Experiments in Fluids*, 61(9):197. https://doi.org/10.1007/s00348-020-03021-6

8. Rajendran LK, Bane SPM, Vlachos PP (2019) Dot tracking methodology for background-oriented schlieren (BOS). *Experiments in Fluids*, 60(11):162. https://doi.org/10.1007/s00348-019-2793-3

9. Rajendran LK, Bane SPM, Vlachos PP (2020) Uncertainty amplification due to density/refractive index gradients in background-oriented schlieren experiments. *Experiments in Fluids*, 61(6):139. https://doi.org/10.1007/s00348-020-02978-8

10. Rajendran LK, Zhang J, Bhattacharya S, Bane SPM, Vlachos PP (2020) Uncertainty quantification in density estimation from background-oriented Schlieren measurements. *Measurement Science and Technology*, 31(5):054002. https://doi.org/10.1088/1361-6501/ab60c8

11. Rajendran L, Zhang J, Bane S, Vlachos P (2020) Uncertainty-based weighted least squares density integration for background-oriented schlieren. *Experiments in Fluids*, 61(11):239. https://doi.org/10.1007/s00348-020-03071-w

12. Raffel M, Willert CE, Scarano F, Kähler CJ, Wereley ST, Kompenhans J (2018) Particle Image Velocimetry, A Practical Guide. https://doi.org/10.1007/978-3-319-68852-7

13. Charonko JJ, Vlachos PP (2013) Estimation of uncertainty bounds for individual particle image velocimetry measurements from cross-correlation peak ratio. *Measurement Science and Technology*, 24(6):065301. https://doi.org/10.1088/0957-0233/24/6/065301

14. Xue Z, Charonko JJ, Vlachos PP (2014) Particle image velocimetry correlation signal-to-noise ratio metrics and measurement uncertainty quantification. *Measurement Science and Technology*, 25(11):115301. https://doi.org/10.1088/0957-0233/25/11/115301

15. Xue Z, Charonko JJ, Vlachos PP (2015) Particle image pattern mutual information and uncertainty estimation for particle image velocimetry. *Measurement Science and Technology*, 26(7):074001. https://doi.org/10.1088/0957-0233/26/7/074001

16. Timmins BH, Wilson BW, Smith BL, Vlachos PP (2012) A method for automatic estimation of instantaneous local uncertainty in particle image velocimetry measurements. *Experiments in Fluids*, 53(4):1133–1147. https://doi.org/10.1007/s00348-012-1341-1

17. Sciacchitano A, Wieneke B, Scarano F (2013) PIV uncertainty quantification by image matching. *Measurement Science and Technology*, 24(4):045302. https://doi.org/10.1088/0957-0233/24/4/045302

18. Wieneke B (2015) PIV uncertainty quantification from correlation statistics. *Measurement Science and Technology*, 26(7)https://doi.org/10.1088/0957-0233/26/7/074002

19. Bhattacharya S, Charonko JJ, Vlachos PP (2018) Particle image velocimetry (PIV) uncertainty quantification using moment of correlation (MC) plane. *Measurement Science and Technology*, 29(11):115301. https://doi.org/10.1088/1361-6501/aadfb4

20. Knapp C, Carter G (1976) The generalized correlation method for estimation of time delay. *IEEE Transactions on Acoustics, Speech, and Signal Processing*, 24(4):320–327. https://doi.org/10.1109/tassp.1976.1162830

21. Wernet MP (2005) Symmetric phase only filtering: a new paradigm for DPIV data processing. *Measurement Science and Technology*, 16(3):601. https://doi.org/10.1088/0957-0233/16/3/001

22. Eckstein AC, Charonko J, Vlachos P (2008) Phase correlation processing for DPIV measurements. *Experiments in Fluids*, 45(3):485–500. https://doi.org/10.1007/s00348-008-0492-6

23. Sciacchitano A, Neal DR, Smith BL, Warner SO, Vlachos PP, Wieneke B, Scarano F (2015) Collaborative framework for PIV uncertainty quantification: comparative assessment of methods. *Measurement Science and Technology*, 26(7):074004. https://doi.org/10.1088/0957-0233/26/7/074004

24. Boomsma A, Bhattacharya S, Troolin D, Pothos S, Vlachos P (2016) A comparative experimental evaluation of uncertainty estimation methods for two-component PIV.

*Measurement Science and Technology*, 27(9):094006. https://doi.org/10.1088/0957-0233/27/9/094006

25. Timmermann A (2013) Forecast Combinations.

26. Bates J, Research C of the (1969) The combination of forecasts. https://doi.org/10.1057/jors.1969.103

27. Clemen RT (1989) Combining forecasts: A review and annotated bibliography. *International Journal of Forecasting*, (5):559–583. https://doi.org/10.1016/0169-2070(89)90012-5

28. Timmermann A (2006) Forecast combinations. *Handbook of economic forecasting*,

29. Marxen M, Sullivan PE, Loewen MR, Jähne B (2000) Comparison of Gaussian particle center estimators and the achievable measurement density for particle tracking velocimetry. *Experiments in Fluids*, 29(2):145–153. https://doi.org/10.1007/s003489900085

30. Cardwell ND, Vlachos PP, Thole KA (2011) A multi-parametric particle-pairing algorithm for particle tracking in single and multiphase flows. *Measurement Science and Technology*, 22(10):105406. https://doi.org/10.1088/0957-0233/22/10/105406

31. Stanislas M, Okamoto K, Kähler CJ, Westerweel J (2005) Main results of the Second International PIV Challenge. *Experiments in Fluids*, 39(2):170–191. https://doi.org/10.1007/s00348-005-0951-2

32. Stanislas M, Okamoto K, Kähler CJ, Westerweel J, Scarano F (2008) Main results of the third international PIV Challenge. *Experiments in Fluids*, 45(1):27–71. https://doi.org/10.1007/s00348-008-0462-z

33. Hubble DO, Vlachos PP, Diller TE (2013) The role of large-scale vortical structures in transient convective heat transfer augmentation. *Journal of Fluid Mechanics*, 718:89–115. https://doi.org/10.1017/jfm.2012.589

34. Kähler CJ, Astarita T, Vlachos PP, Sakakibara J, Hain R, Discetti S, Foy RL, Cierpka C (2016) Main results of the 4th International PIV Challenge. *Experiments in Fluids*, 57(6):97. https://doi.org/10.1007/s00348-016-2173-1

35. (n.d.) PRANA: PIV Research and Analysis. https://github.com/aether-lab/prana/

36. Eckstein A, Vlachos PP (2009) Assessment of advanced windowing techniques for digital particle image velocimetry (DPIV). *Measurement Science and Technology*, 20(7):075402. https://doi.org/10.1088/0957-0233/20/7/075402

37. Raffel M, Willert CE, Wereley ST, Kompenhans J (2007) Particle Image Velocimetry, A Practical Guide. https://doi.org/10.1007/978-3-540-72308-0

38. Westerweel J, Dabiri D, Gharib M (1997) The effect of a discrete window offset on the accuracy of cross-correlation analysis of digital PIV recordings. *Experiments in Fluids*, 23(1):20–28. https://doi.org/10.1007/s003480050082

39. Westerweel J (2000) Theoretical analysis of the measurement precision in particle image velocimetry. *Experiments in Fluids*, 29(Suppl 1):S003-S012. https://doi.org/10.1007/s003480070002

40. Westerweel J (2008) On velocity gradients in PIV interrogation. *Experiments in Fluids*, 44(5)https://doi.org/10.1007/s00348-007-0439-3

41. Wieneke B (2005) Stereo-PIV using self-calibration on particle images. *Experiments in Fluids*, 39(2):267–280. https://doi.org/10.1007/s00348-005-0962-z

# 7 Appendix A: Methods for perturbing the particle images

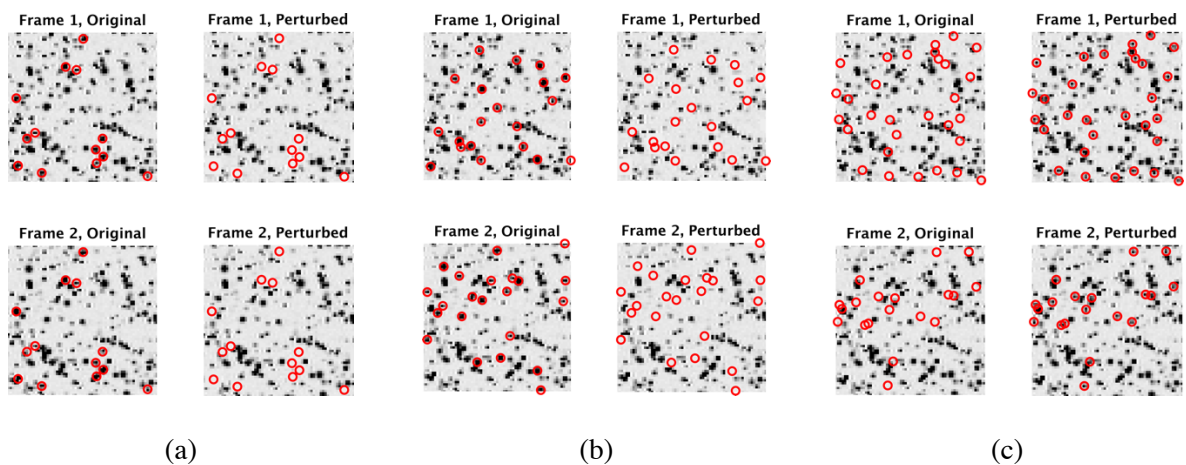

(a)　　　　　　　　　　　　(b)　　　　　　　　　　　　(c)

**Figure 12.** Illustration of methods used to perturb the particle images. (a) Removing paired particles, (b) Removing unpaired particles, (c) Adding unpaired particles.

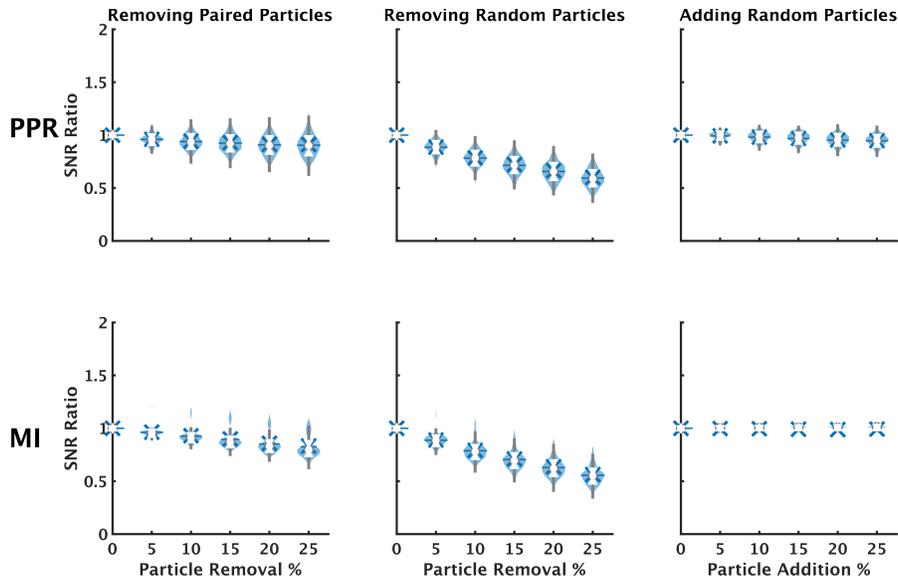

**Figure 13.** Effect of particle perturbation on correlation plane SNR metrics for the three methods.

# 8 Appendix B: Planar Uncertainty Distributions for individual datasets

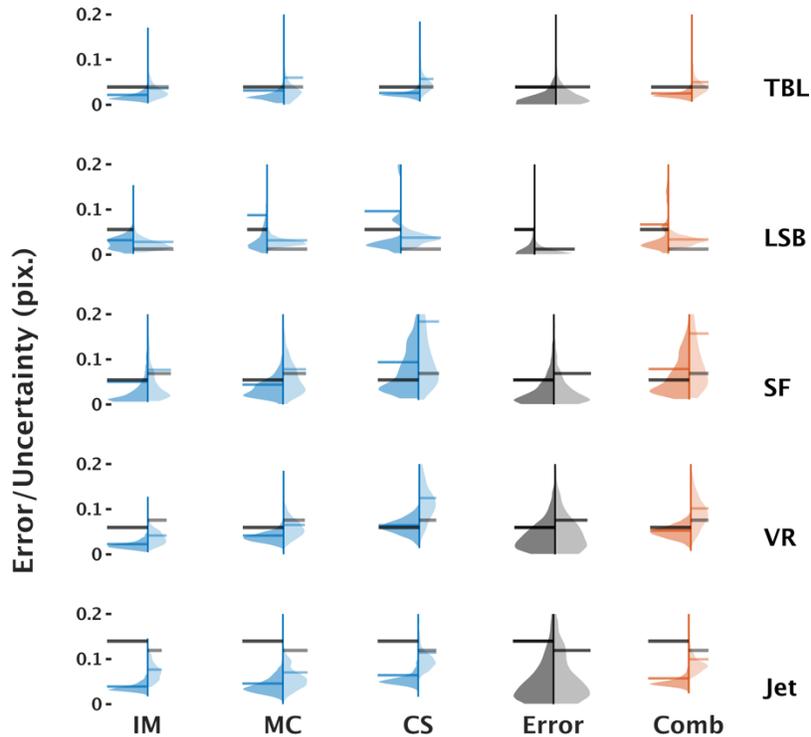

**Figure 14.** Error and uncertainty distributions for each of the planar PIV datasets. For each violin plot, the left (darker) and right (lighter) halves correspond to the results for WS1 and WS2 processing, respectively.

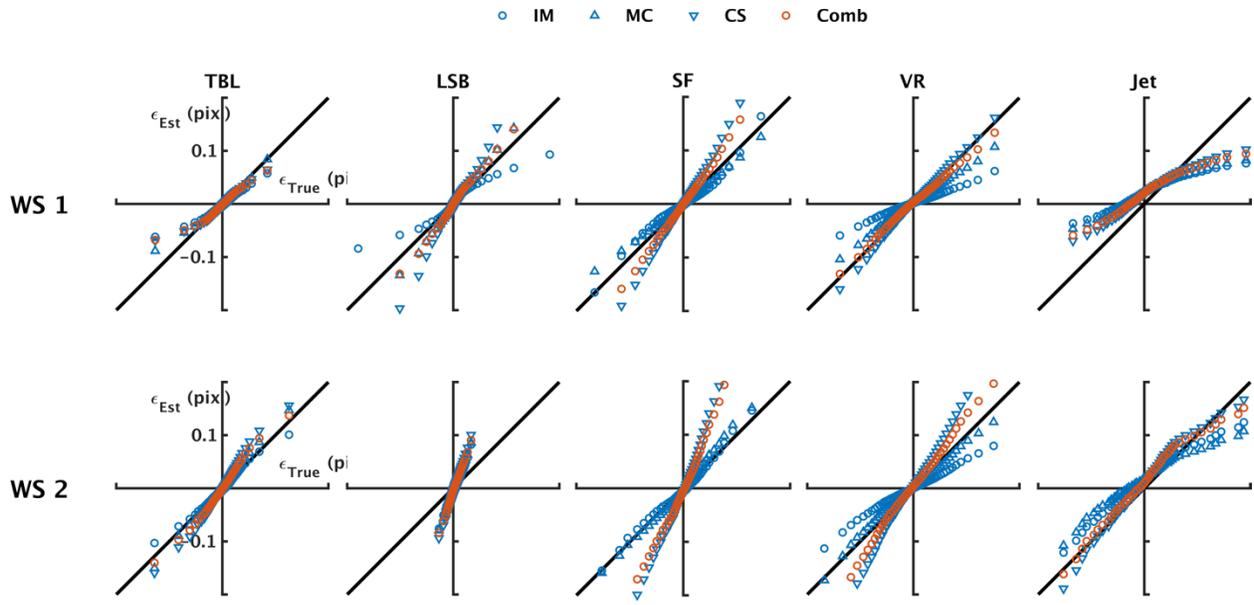

**Figure 15.** Quantile-Quantile comparisons of the true and estimated error distributions based on the individual and combined uncertainty estimates for each planar PIV dataset.